

%
\catcode`@=11 
%
%
%

\font\fourteenrm=cmr10 scaled\magstep2
\font\twelverm=cmr10 scaled\magstep1
\font\ninerm=cmr9            \font\sixrm=cmr6

\font\fourteenbf=cmbx10 scaled\magstep2
\font\twelvebf=cmbx10 scaled\magstep1
\font\ninebf=cmbx9            \font\sixbf=cmbx6
\font\seventeeni=cmmi10 scaled\magstep3     \skewchar\seventeeni='177
\font\fourteeni=cmmi10 scaled\magstep2      \skewchar\fourteeni='177
\font\twelvei=cmmi10 scaled\magstep1        \skewchar\twelvei='177
\font\ninei=cmmi9                           \skewchar\ninei='177
\font\sixi=cmmi6                            \skewchar\sixi='177
\font\seventeensy=cmsy10 scaled\magstep3    \skewchar\seventeensy='60
\font\fourteensy=cmsy10 scaled\magstep2     \skewchar\fourteensy='60
\font\twelvesy=cmsy10 scaled\magstep1       \skewchar\twelvesy='60
\font\ninesy=cmsy9                          \skewchar\ninesy='60
\font\sixsy=cmsy6                           \skewchar\sixsy='60

\font\fourteenex=cmex10 scaled\magstep2
\font\twelveex=cmex10 scaled\magstep1

\font\fourteensl=cmsl10 scaled\magstep2
\font\twelvesl=cmsl10 scaled\magstep1
\font\ninesl=cmsl9

\font\fourteenit=cmti10 scaled\magstep2
\font\twelveit=cmti10 scaled\magstep1
\font\twelvett=cmtt10 scaled\magstep1
\font\twelvecp=cmcsc10 scaled\magstep1
\font\tencp=cmcsc10
\newfam\cpfam
%
%
\newcount\f@ntkey            \f@ntkey=0
\def\samef@nt{\relax \ifcase\f@ntkey \rm \or\oldstyle \or\or
         \or\it \or\sl \or\bf \or\tt \or\caps \fi }
\def\fourteenpoint{\relax
    \textfont0=\fourteenrm          \scriptfont0=\tenrm
    \scriptscriptfont0=\sevenrm
     \def\rm{\fam0 \fourteenrm \f@ntkey=0 }\relax
    \textfont1=\fourteeni           \scriptfont1=\teni
    \scriptscriptfont1=\seveni
     \def\oldstyle{\fam1 \fourteeni\f@ntkey=1 }\relax
    \textfont2=\fourteensy          \scriptfont2=\tensy
    \scriptscriptfont2=\sevensy
    \textfont3=\fourteenex     \scriptfont3=\fourteenex
    \scriptscriptfont3=\fourteenex
    \def\it{\fam\itfam \fourteenit\f@ntkey=4 }\textfont\itfam=\fourteenit
    \def\sl{\fam\slfam \fourteensl\f@ntkey=5 }\textfont\slfam=\fourteensl
    \scriptfont\slfam=\tensl
    \def\bf{\fam\bffam \fourteenbf\f@ntkey=6 }\textfont\bffam=\fourteenbf
    \scriptfont\bffam=\tenbf     \scriptscriptfont\bffam=\sevenbf
    \def\tt{\fam\ttfam \twelvett \f@ntkey=7 }\textfont\ttfam=\twelvett
    \h@big=11.9\p@{} \h@Big=16.1\p@{} \h@bigg=20.3\p@{} \h@Bigg=24.5\p@{}
    \def\caps{\fam\cpfam \twelvecp \f@ntkey=8 }\textfont\cpfam=\twelvecp
    \setbox\strutbox=\hbox{\vrule height 12pt depth 5pt width\z@}
    \samef@nt}
\def\twelvepoint{\relax
    \textfont0=\twelverm          \scriptfont0=\ninerm
    \scriptscriptfont0=\sixrm
     \def\rm{\fam0 \twelverm \f@ntkey=0 }\relax
    \textfont1=\twelvei           \scriptfont1=\ninei
    \scriptscriptfont1=\sixi
     \def\oldstyle{\fam1 \twelvei\f@ntkey=1 }\relax
    \textfont2=\twelvesy          \scriptfont2=\ninesy
    \scriptscriptfont2=\sixsy
    \textfont3=\twelveex          \scriptfont3=\twelveex
    \scriptscriptfont3=\twelveex
    \def\it{\fam\itfam \twelveit \f@ntkey=4 }\textfont\itfam=\twelveit
    \def\sl{\fam\slfam \twelvesl \f@ntkey=5 }\textfont\slfam=\twelvesl
    \scriptfont\slfam=\ninesl
    \def\bf{\fam\bffam \twelvebf \f@ntkey=6 }\textfont\bffam=\twelvebf
    \scriptfont\bffam=\ninebf     \scriptscriptfont\bffam=\sixbf
    \def\tt{\fam\ttfam \twelvett \f@ntkey=7 }\textfont\ttfam=\twelvett
    \h@big=10.2\p@{}
    \h@Big=13.8\p@{}
    \h@bigg=17.4\p@{}
    \h@Bigg=21.0\p@{}
    \def\caps{\fam\cpfam \twelvecp \f@ntkey=8 }\textfont\cpfam=\twelvecp
    \setbox\strutbox=\hbox{\vrule height 10pt depth 4pt width\z@}
    \samef@nt}
\def\tenpoint{\relax
    \textfont0=\tenrm          \scriptfont0=\sevenrm
    \scriptscriptfont0=\fiverm
    \def\rm{\fam0 \tenrm \f@ntkey=0 }\relax
    \textfont1=\teni           \scriptfont1=\seveni
    \scriptscriptfont1=\fivei
    \def\oldstyle{\fam1 \teni \f@ntkey=1 }\relax
    \textfont2=\tensy          \scriptfont2=\sevensy
    \scriptscriptfont2=\fivesy
    \textfont3=\tenex          \scriptfont3=\tenex
    \scriptscriptfont3=\tenex
    \def\it{\fam\itfam \tenit \f@ntkey=4 }\textfont\itfam=\tenit
    \def\sl{\fam\slfam \tensl \f@ntkey=5 }\textfont\slfam=\tensl
    \def\bf{\fam\bffam \tenbf \f@ntkey=6 }\textfont\bffam=\tenbf
    \scriptfont\bffam=\sevenbf     \scriptscriptfont\bffam=\fivebf
    \def\tt{\fam\ttfam \tentt \f@ntkey=7 }\textfont\ttfam=\tentt
    \def\caps{\fam\cpfam \tencp \f@ntkey=8 }\textfont\cpfam=\tencp
    \setbox\strutbox=\hbox{\vrule height 8.5pt depth 3.5pt width\z@}
    \samef@nt}
%
%
%
%
\newdimen\h@big  \h@big=8.5\p@
\newdimen\h@Big  \h@Big=11.5\p@
\newdimen\h@bigg  \h@bigg=14.5\p@
\newdimen\h@Bigg  \h@Bigg=17.5\p@
\def\big#1{{\hbox{$\left#1\vbox to\h@big{}\right.\n@space$}}}
\def\Big#1{{\hbox{$\left#1\vbox to\h@Big{}\right.\n@space$}}}
\def\bigg#1{{\hbox{$\left#1\vbox to\h@bigg{}\right.\n@space$}}}
\def\Bigg#1{{\hbox{$\left#1\vbox to\h@Bigg{}\right.\n@space$}}}
%
%
%
\normalbaselineskip = 20pt plus 0.2pt minus 0.1pt
\normallineskip = 1.5pt plus 0.1pt minus 0.1pt
\normallineskiplimit = 1.5pt
\newskip\normaldisplayskip
\normaldisplayskip = 20pt plus 5pt minus 10pt
\newskip\normaldispshortskip
\normaldispshortskip = 6pt plus 5pt
\newskip\normalparskip
\normalparskip = 6pt plus 2pt minus 1pt
\newskip\skipregister
\skipregister = 5pt plus 2pt minus 1.5pt
\newif\ifsingl@    \newif\ifdoubl@
\newif\iftwelv@    \twelv@true
\def\singlespace{\singl@true\doubl@false\spaces@t}
\def\doublespace{\singl@false\doubl@true\spaces@t}
\def\normalspace{\singl@false\doubl@false\spaces@t}
\def\Tenpoint{\tenpoint\twelv@false\spaces@t}
\def\Twelvepoint{\twelvepoint\twelv@true\spaces@t}
\def\spaces@t{\relax%
 \iftwelv@ \ifsingl@\subspaces@t3:4;\else\subspaces@t1:1;\fi%
 \else \ifsingl@\subspaces@t3:5;\else\subspaces@t4:5;\fi \fi%
 \ifdoubl@ \multiply\baselineskip by 5%
 \divide\baselineskip by 4 \fi \unskip}
\def\subspaces@t#1:#2;{
      \baselineskip = \normalbaselineskip
      \multiply\baselineskip by #1 \divide\baselineskip by #2
      \lineskip = \normallineskip
      \multiply\lineskip by #1 \divide\lineskip by #2
      \lineskiplimit = \normallineskiplimit
      \multiply\lineskiplimit by #1 \divide\lineskiplimit by #2
      \parskip = \normalparskip
      \multiply\parskip by #1 \divide\parskip by #2
      \abovedisplayskip = \normaldisplayskip
      \multiply\abovedisplayskip by #1 \divide\abovedisplayskip by #2
      \belowdisplayskip = \abovedisplayskip
      \abovedisplayshortskip = \normaldispshortskip
      \multiply\abovedisplayshortskip by #1
        \divide\abovedisplayshortskip by #2
      \belowdisplayshortskip = \abovedisplayshortskip
      \advance\belowdisplayshortskip by \belowdisplayskip
      \divide\belowdisplayshortskip by 2
      \smallskipamount = \skipregister
      \multiply\smallskipamount by #1 \divide\smallskipamount by #2
      \medskipamount = \smallskipamount \multiply\medskipamount by 2
      \bigskipamount = \smallskipamount \multiply\bigskipamount by 4 }
\def\normalbaselines{ \baselineskip=\normalbaselineskip
   \lineskip=\normallineskip \lineskiplimit=\normallineskip
   \iftwelv@\else \multiply\baselineskip by 4 \divide\baselineskip by 5
     \multiply\lineskiplimit by 4 \divide\lineskiplimit by 5
     \multiply\lineskip by 4 \divide\lineskip by 5 \fi }
\Twelvepoint  
\interlinepenalty=50
\interfootnotelinepenalty=5000
\predisplaypenalty=9000
\postdisplaypenalty=500
\hfuzz=1pt
\vfuzz=0.2pt
%
%
%
\def\pagecontents{
   \ifvoid\topins\else\unvbox\topins\vskip\skip\topins\fi
   \dimen@ = \dp255 \unvbox255
   \ifvoid\footins\else\vskip\skip\footins\footrule\unvbox\footins\fi
   \ifr@ggedbottom \kern-\dimen@ \vfil \fi }
\def\makeheadline{\vbox to 0pt{ \skip@=\topskip
      \advance\skip@ by -12pt \advance\skip@ by -2\normalbaselineskip
      \vskip\skip@ \line{\vbox to 12pt{}\the\headline} \vss
      }\nointerlineskip}
\def\makefootline{\baselineskip = 1.5\normalbaselineskip
                 \line{\the\footline}}
\newif\iffrontpage
\newif\ifletterstyle
\newif\ifp@genum
\def\nopagenumbers{\p@genumfalse}
\def\pagenumbers{\p@genumtrue}
\pagenumbers
\newtoks\paperheadline
\newtoks\letterheadline
\newtoks\letterfrontheadline
\newtoks\lettermainheadline
\newtoks\paperfootline
\newtoks\letterfootline
\newtoks\date
\footline={\ifletterstyle\the\letterfootline\else\the\paperfootline\fi}
\paperfootline={\hss\iffrontpage\else\ifp@genum\tenrm\folio\hss\fi\fi}
\letterfootline={\hfil}
\headline={\ifletterstyle\the\letterheadline\else\the\paperheadline\fi}
\paperheadline={\hfil}
\letterheadline{\iffrontpage\the\letterfrontheadline
     \else\the\lettermainheadline\fi}
\lettermainheadline={\rm\ifp@genum page \ \folio\fi\hfil\the\date}
\def\monthname{\relax\ifcase\month 0/\or January\or February\or
   March\or April\or May\or June\or July\or August\or September\or
   October\or November\or December\else\number\month/\fi}
\date={\monthname\ \number\day, \number\year}
\countdef\pagenumber=1  \pagenumber=1
\def\advancepageno{\global\advance\pageno by 1
   \ifnum\pagenumber<0 \global\advance\pagenumber by -1
    \else\global\advance\pagenumber by 1 \fi \global\frontpagefalse }
\def\folio{\ifnum\pagenumber<0 \romannumeral-\pagenumber
           \else \number\pagenumber \fi }
\def\footrule{\dimen@=\prevdepth\nointerlineskip
   \vbox to 0pt{\vskip -0.25\baselineskip \hrule width 0.35\hsize \vss}
   \prevdepth=\dimen@ }
\newtoks\foottokens
\foottokens={\Tenpoint\singlespace}
\newdimen\footindent
\footindent=24pt
\def\vfootnote#1{\insert\footins\bgroup  \the\foottokens
   \interlinepenalty=\interfootnotelinepenalty \floatingpenalty=20000
   \splittopskip=\ht\strutbox \boxmaxdepth=\dp\strutbox
   \leftskip=\footindent \rightskip=\z@skip
   \parindent=0.5\footindent \parfillskip=0pt plus 1fil
   \spaceskip=\z@skip \xspaceskip=\z@skip
   \Textindent{$ #1 $}\footstrut\futurelet\next\fo@t}
\def\Textindent#1{\noindent\llap{#1\enspace}\ignorespaces}
\def\footnote#1{\attach{#1}\vfootnote{#1}}

\def\foot{\attach\footsymbolgen\vfootnote{\footsymbol}}
\let\footsymbol=\star
\newcount\lastf@@t           \lastf@@t=-1
\newcount\footsymbolcount    \footsymbolcount=0
\newif\ifPhysRev
\def\footsymbolgen{\relax \ifPhysRev \iffrontpage \NPsymbolgen\else
      \PRsymbolgen\fi \else \NPsymbolgen\fi
   \global\lastf@@t=\pageno \footsymbol }
\def\NPsymbolgen{\ifnum\footsymbolcount<0 \global\footsymbolcount=0\fi
   {\iffrontpage \else \advance\lastf@@t by 1 \fi
    \ifnum\lastf@@t<\pageno \global\footsymbolcount=0
     \else \global\advance\footsymbolcount by 1 \fi }
   \ifcase\footsymbolcount \fd@f\star\or \fd@f\dagger\or \fd@f\ast\or
    \fd@f\ddagger\or \fd@f\natural\or \fd@f\diamond\or \fd@f\bullet\or
    \fd@f\nabla\else \fd@f\dagger\global\footsymbolcount=0 \fi }
\def\fd@f#1{\xdef\footsymbol{#1}}
\def\PRsymbolgen{\ifnum\footsymbolcount>0 \global\footsymbolcount=0\fi
      \global\advance\footsymbolcount by -1
      \xdef\footsymbol{\sharp\number-\footsymbolcount} }
\def\space@ver#1{\let\@sf=\empty \ifmmode #1\else \ifhmode
   \edef\@sf{\spacefactor=\the\spacefactor}\unskip${}#1$\relax\fi\fi}
\def\attach#1{\space@ver{\strut^{\mkern 2mu #1} }\@sf\ }
%
%
%
\newcount\chapternumber      \chapternumber=0
\newcount\sectionnumber      \sectionnumber=0
\newcount\equanumber         \equanumber=0
\let\chapterlabel=0
\newtoks\chapterstyle        \chapterstyle={\Number}
\newskip\chapterskip         \chapterskip=\bigskipamount
\newskip\sectionskip         \sectionskip=\medskipamount
\newskip\headskip            \headskip=8pt plus 3pt minus 3pt
\newdimen\chapterminspace    \chapterminspace=15pc
\newdimen\sectionminspace    \sectionminspace=10pc
\newdimen\referenceminspace  \referenceminspace=25pc
\def\chapterreset{\global\advance\chapternumber by 1
   \ifnum\the\equanumber<0 \else\global\equanumber=0\fi
   \sectionnumber=0 \makel@bel}
\def\makel@bel{\xdef\chapterlabel{%
\the\chapterstyle{\the\chapternumber}.}}
\def\sectionlabel{\number\sectionnumber \quad }
\def\alphabetic#1{\count255='140 \advance\count255 by #1\char\count255}
\def\Alphabetic#1{\count255='100 \advance\count255 by #1\char\count255}
\def\Roman#1{\uppercase\expandafter{\romannumeral #1}}
\def\roman#1{\romannumeral #1}
\def\Number#1{\number #1}
\def\unnumberedchapters{\let\makel@bel=\relax \let\chapterlabel=\relax
\let\sectionlabel=\relax \equanumber=-1 }
\def\titlestyle#1{\par\begingroup \interlinepenalty=9999
     \leftskip=0.02\hsize plus 0.23\hsize minus 0.02\hsize
     \rightskip=\leftskip \parfillskip=0pt
     \hyphenpenalty=9000 \exhyphenpenalty=9000
     \tolerance=9999 \pretolerance=9000
     \spaceskip=0.333em \xspaceskip=0.5em
     \iftwelv@\fourteenpoint\else\twelvepoint\fi
   \noindent #1\par\endgroup }
\def\spacecheck#1{\dimen@=\pagegoal\advance\dimen@ by -\pagetotal
   \ifdim\dimen@<#1 \ifdim\dimen@>0pt \vfil\break \fi\fi}
\def\chapter#1{\par \penalty-300 \vskip\chapterskip
   \spacecheck\chapterminspace
   \chapterreset \titlestyle{\chapterlabel \ #1}
   \nobreak\vskip\headskip \penalty 30000
   \wlog{\string\chapter\ \chapterlabel} }

\def\section#1{\par \ifnum\the\lastpenalty=30000\else
   \penalty-200\vskip\sectionskip \spacecheck\sectionminspace\fi
   \wlog{\string\section\ \chapterlabel \the\sectionnumber}
   \global\advance\sectionnumber by 1  \noindent
   {\caps\enspace\chapterlabel \sectionlabel #1}\par
   \nobreak\vskip\headskip \penalty 30000 }
\def\subsection#1{\par
   \ifnum\the\lastpenalty=30000\else \penalty-100\smallskip \fi
   \noindent\undertext{#1}\enspace \vadjust{\penalty5000}}

\def\undertext#1{\vtop{\hbox{#1}\kern 1pt \hrule}}
\def\APPENDIX#1#2{\par\penalty-300\vskip\chapterskip
   \spacecheck\chapterminspace \chapterreset \xdef\chapterlabel{#1}
   \titlestyle{APPENDIX #2} \nobreak\vskip\headskip \penalty 30000
   \wlog{\string\Appendix\ \chapterlabel} }
\def\Appendix#1{\APPENDIX{#1}{#1}}
\def\appendix{\APPENDIX{A}{}}
%
%
%
\def\eqname#1{\relax \ifnum\the\equanumber<0
     \xdef#1{{\rm(\number-\equanumber)}}\global\advance\equanumber by -1
    \else \global\advance\equanumber by 1
      \xdef#1{{\rm(\chapterlabel \number\equanumber)}} \fi}
\def\eqinsert#1{\noalign{\dimen@=\prevdepth \nointerlineskip
   \setbox0=\hbox to\displaywidth{\hfil #1}
   \vbox to 0pt{\vss\hbox{$\!\box0\!$}\kern-0.5\baselineskip}
   \prevdepth=\dimen@}}
%

%

%

%
%
\def\GENITEM#1;#2{\par \hangafter=0 \hangindent=#1
    \Textindent{$ #2 $}\ignorespaces}
\outer\def\newitem#1=#2;{\gdef#1{\GENITEM #2;}}
\newdimen\itemsize                \itemsize=30pt
\newitem\item=1\itemsize;
\newitem\sitem=1.75\itemsize;     
\newitem\ssitem=2.5\itemsize;     
\outer\def\newlist#1=#2&#3&#4;{\toks0={#2}\toks1={#3}%
   \count255=\escapechar \escapechar=-1
   \alloc@0\list\countdef\insc@unt\listcount     \listcount=0
   \edef#1{\par
      \countdef\listcount=\the\allocationnumber
      \advance\listcount by 1
      \hangafter=0 \hangindent=#4
      \Textindent{\the\toks0{\listcount}\the\toks1}}
   \expandafter\expandafter\expandafter
    \edef\c@t#1{begin}{\par
      \countdef\listcount=\the\allocationnumber \listcount=1
      \hangafter=0 \hangindent=#4
      \Textindent{\the\toks0{\listcount}\the\toks1}}
   \expandafter\expandafter\expandafter
    \edef\c@t#1{con}{\par \hangafter=0 \hangindent=#4 \noindent}
   \escapechar=\count255}
\def\c@t#1#2{\csname\string#1#2\endcsname}
\newlist\point=\Number&.&1.0\itemsize;
\newlist\subpoint=(\alphabetic&)&1.75\itemsize;
\newlist\subsubpoint=(\roman&)&2.5\itemsize;
\newlist\cpoint=\Roman&.&1.0\itemsize;
%

%
%
%
\newcount\referencecount     \referencecount=0
\newif\ifreferenceopen       \newwrite\referencewrite
\newtoks\rw@toks
\def\NPrefmark#1{\attach{\scriptscriptstyle [ #1 ] }}
\let\PRrefmark=\attach
\def\CErefmark#1{\attach{\scriptstyle  #1 ) }}
\def\refmark#1{\relax\ifPhysRev\PRrefmark{#1}\else\NPrefmark{#1}\fi}
\def\crefmark#1{\relax\CErefmark{#1}}
\def\refend{\refmark{\number\referencecount}}
\newcount\lastrefsbegincount \lastrefsbegincount=0
\def\refsend{\refmark{\count255=\referencecount
   \advance\count255 by-\lastrefsbegincount
   \ifcase\count255 \number\referencecount
   \or \number\lastrefsbegincount,\number\referencecount
   \else \number\lastrefsbegincount-\number\referencecount \fi}}
\def\crefsend{\crefmark{\count255=\referencecount
   \advance\count255 by-\lastrefsbegincount
   \ifcase\count255 \number\referencecount
   \or \number\lastrefsbegincount,\number\referencecount
   \else \number\lastrefsbegincount-\number\referencecount \fi}}
\def\refch@ck{\chardef\rw@write=\referencewrite
   \ifreferenceopen \else \referenceopentrue
   \immediate\openout\referencewrite=referenc.texauxil \fi}
%
{\catcode`\^^M=\active 
  \gdef\obeyendofline{\catcode`\^^M\active \let^^M\ }}%
%
{\catcode`\^^M=\active 
  \gdef\ignoreendofline{\catcode`\^^M=5}}
{\obeyendofline\gdef\rw@start#1{\def\t@st{#1} \ifx\t@st\blankend%
\endgroup \@sf \relax \else \ifx\t@st\bl@nkend \endgroup \@sf \relax%
\else \rw@begin#1
\backtotext
\fi \fi } }
{\obeyendofline\gdef\rw@begin#1
{\def\n@xt{#1}\rw@toks={#1}\relax%
\rw@next}}
\def\blankend{}
{\obeylines\gdef\bl@nkend{
}}
\newif\iffirstrefline  \firstreflinetrue
\def\rwr@teswitch{\ifx\n@xt\blankend \let\n@xt=\rw@begin %
 \else\iffirstrefline \global\firstreflinefalse%
\immediate\write\rw@write{\noexpand\obeyendofline \the\rw@toks}%
\let\n@xt=\rw@begin%
      \else\ifx\n@xt\rw@@d \def\n@xt{\immediate\write\rw@write{%
        \noexpand\ignoreendofline}\endgroup \@sf}%
             \else \immediate\write\rw@write{\the\rw@toks}%
             \let\n@xt=\rw@begin\fi\fi \fi}
\def\rw@next{\rwr@teswitch\n@xt}
\def\rw@@d{\backtotext} \let\rw@end=\relax
\let\backtotext=\relax

\newdimen\refindent     \refindent=30pt
\def\refitem#1{\par \hangafter=0 \hangindent=\refindent \Textindent{#1}}
\def\REFNUM#1{\space@ver{}\refch@ck \firstreflinetrue%
 \global\advance\referencecount by 1 \xdef#1{\the\referencecount}}
\def\refnum#1{\space@ver{}\refch@ck \firstreflinetrue%
 \global\advance\referencecount by 1 \xdef#1{\the\referencecount}\refend}

\def\REF#1{\REFNUM#1%
 \immediate\write\referencewrite{%
 \noexpand\refitem{#1.}}%
\begingroup\obeyendofline\rw@start}
\def\ref{\refnum\?%
 \immediate\write\referencewrite{\noexpand\refitem{\?.}}%
\begingroup\obeyendofline\rw@start}
\def\Ref#1{\refnum#1%
 \immediate\write\referencewrite{\noexpand\refitem{#1.}}%
\begingroup\obeyendofline\rw@start}
\def\REFS#1{\REFNUM#1\global\lastrefsbegincount=\referencecount
\immediate\write\referencewrite{\noexpand\refitem{#1.}}%
\begingroup\obeyendofline\rw@start}
\newcount\figurecount     \figurecount=0
\newif\iffigureopen       \newwrite\figurewrite
\def\figch@ck{\chardef\rw@write=\figurewrite \iffigureopen\else
   \immediate\openout\figurewrite=figures.texauxil
   \figureopentrue\fi}
\def\FIGNUM#1{\space@ver{}\figch@ck \firstreflinetrue%
 \global\advance\figurecount by 1 \xdef#1{\the\figurecount}}
\def\FIG#1{\FIGNUM#1
   \immediate\write\figurewrite{\noexpand\refitem{#1.}}%
   \begingroup\obeyendofline\rw@start}
\def\fig{\FIGNUM\? fig.~\?%
\immediate\write\figurewrite{\noexpand\refitem{\?.}}%
\begingroup\obeyendofline\rw@start}
\def\figure{\FIGNUM\? figure~\?
   \immediate\write\figurewrite{\noexpand\refitem{\?.}}%
   \begingroup\obeyendofline\rw@start}
\def\Fig{\FIGNUM\? Fig.~\?%
\immediate\write\figurewrite{\noexpand\refitem{\?.}}%
\begingroup\obeyendofline\rw@start}
\def\Figure{\FIGNUM\? Figure~\?%
\immediate\write\figurewrite{\noexpand\refitem{\?.}}%
\begingroup\obeyendofline\rw@start}
\newcount\tablecount     \tablecount=0
\newif\iftableopen       \newwrite\tablewrite
\def\tabch@ck{\chardef\rw@write=\tablewrite \iftableopen\else
   \immediate\openout\tablewrite=tables.texauxil
   \tableopentrue\fi}
\def\TABNUM#1{\space@ver{}\tabch@ck \firstreflinetrue%
 \global\advance\tablecount by 1 \xdef#1{\the\tablecount}}
\def\TABLE#1{\TABNUM#1
   \immediate\write\tablewrite{\noexpand\refitem{#1.}}%
   \begingroup\obeyendofline\rw@start}
\def\Table{\TABNUM\? Table~\?%
\immediate\write\tablewrite{\noexpand\refitem{\?.}}%
\begingroup\obeyendofline\rw@start}
%

%
%
%
\def\masterreset{\global\pagenumber=1 \global\chapternumber=0
   \ifnum\the\equanumber<0\else \global\equanumber=0\fi
   \global\sectionnumber=0
   \global\referencecount=0 \global\figurecount=0 \global\tablecount=0 }
\def\FRONTPAGE{\ifvoid255\else\vfill\penalty-2000\fi
      \masterreset\global\frontpagetrue
      \global\lastf@@t=0 \global\footsymbolcount=0}

\def\paperstyle{\letterstylefalse\normalspace\papersize}
\def\letterstyle{\letterstyletrue\singlespace\lettersize}
\def\papersize{\hsize=35pc\vsize=48pc\hoffset=1pc\voffset=6pc
               \skip\footins=\bigskipamount}
\def\lettersize{\hsize=6.5in\vsize=8.5in\hoffset=0in\voffset=1in
   \skip\footins=\smallskipamount \multiply\skip\footins by 3 }
\paperstyle   
%
%
\def\MEMO{\letterstyle\FRONTPAGE \letterfrontheadline={\hfil}
    \line{\quad\fourteenrm FNAL MEMORANDUM\hfil\twelverm\the\date\quad}
    \medskip \memod@f}

\def\memit@m#1{\smallskip \hangafter=0 \hangindent=1in
      \Textindent{\caps #1}}
\def\memod@f{\xdef\to{\memit@m{To:}}\xdef\from{\memit@m{From:}}%
     \xdef\topic{\memit@m{Topic:}}\xdef\subject{\memit@m{Subject:}}%
     \xdef\rule{\bigskip\hrule height 1pt\bigskip}}
\memod@f
\newskip\lettertopfil
\lettertopfil = 0pt plus 1.5in minus 0pt
\newskip\letterbottomfil
\letterbottomfil = 0pt plus 2.3in minus 0pt
\newskip\spskip \setbox0\hbox{\ } \spskip=-1\wd0
\def\addressee#1{\medskip\rightline{\the\date\hskip 30pt} \bigskip
   \vskip\lettertopfil
   \ialign to\hsize{\strut ##\hfil\tabskip 0pt plus \hsize \cr #1\crcr}
   \medskip\noindent\hskip\spskip}
\newskip\signatureskip       \signatureskip=40pt
\def\signed#1{\par \penalty 9000 \bigskip \dt@pfalse
  \everycr={\noalign{\ifdt@p\vskip\signatureskip\global\dt@pfalse\fi}}
  \setbox0=\vbox{\singlespace \halign{\tabskip 0pt \strut ##\hfil\cr
   \noalign{\global\dt@ptrue}#1\crcr}}
  \line{\hskip 0.5\hsize minus 0.5\hsize \box0\hfil} \medskip }

\def\endletter{\ifnum\pagenumber=1 \vskip\letterbottomfil\supereject
\else \vfil\supereject \fi}
\newbox\letterb@x
\def\lettertext{\par\unvcopy\letterb@x\par}
\def\multiletter{\setbox\letterb@x=\vbox\bgroup
      \everypar{\vrule height 1\baselineskip depth 0pt width 0pt }
      \singlespace \topskip=\baselineskip }
\def\letterend{\par\egroup}
%
%
%
\newskip\frontpageskip
\newtoks\pubtype
\newtoks\Pubnum
\newtoks\pubnum
\newif\ifp@bblock  \p@bblocktrue
\def\PH@SR@V{\doubl@true \baselineskip=24.1pt plus 0.2pt minus 0.1pt
             \parskip= 3pt plus 2pt minus 1pt }
\def\PHYSREV{\paperstyle\PhysRevtrue\PH@SR@V}
\def\titlepage{\FRONTPAGE\paperstyle\ifPhysRev\PH@SR@V\fi
   \ifp@bblock\p@bblock\fi}
\def\nopubblock{\p@bblockfalse}

\frontpageskip=1\medskipamount plus .5fil
\pubtype={\tensl Preliminary Version}
\pubnum={0000}
\def\p@bblock{\begingroup \tabskip=\hsize minus \hsize
   \baselineskip=1.5\ht\strutbox \topspace-2\baselineskip
   \halign to\hsize{\strut ##\hfil\tabskip=0pt\crcr
   \the\Pubnum\cr \the\date\cr}\endgroup}

%
\def\title#1{\vskip\frontpageskip \titlestyle{#1} \vskip\headskip }
\def\author#1{\vskip\frontpageskip\titlestyle{\twelvecp #1}\nobreak}

\def\address#1{\par\kern 5pt\titlestyle{\twelvepoint\it #1}}
\def\andaddress{\par\kern 5pt \centerline{\sl and} \address}

\def\abstract{\vskip\frontpageskip\centerline{\fourteenrm ABSTRACT}
              \vskip\headskip }

%
%
%

\def\\{\relax\ifmmode\backslash\else$\backslash$\fi}
\def\globaleqnumbers{\relax\ifnum\the\equanumber<0%
\else\global\equanumber=-1\fi}

\def\journal#1&#2(#3){\unskip, \sl #1~\bf #2 \rm (19#3) }

\def\topspace{\hrule height 0pt depth 0pt \vskip}

\def\VEV#1{\left\langle #1\right\rangle}

\let\int=\intop         
\def\prop{\mathrel{{\mathchoice{\pr@p\scriptstyle}{\pr@p\scriptstyle}{
                \pr@p\scriptscriptstyle}{\pr@p\scriptscriptstyle} }}}
\def\pr@p#1{\setbox0=\hbox{$\cal #1 \char'103$}
   \hbox{$\cal #1 \char'117$\kern-.4\wd0\box0}}
\def\lsim{\mathrel{\mathpalette\@versim<}}
\def\gsim{\mathrel{\mathpalette\@versim>}}
\def\@versim#1#2{\lower0.2ex\vbox{\baselineskip\z@skip\lineskip\z@skip
  \lineskiplimit\z@\ialign{$\m@th#1\hfil##\hfil$\crcr#2\crcr\sim\crcr}}}
\def\leftrightarrowfill{$\m@th \mathord- \mkern-6mu
	\cleaders\hbox{$\mkern-2mu \mathord- \mkern-2mu$}\hfil
	\mkern-6mu \mathord\leftrightarrow$}
\def\lrover#1{\vbox{\ialign{##\crcr
	\leftrightarrowfill\crcr\noalign{\kern-1pt\nointerlineskip}
	$\hfil\displaystyle{#1}\hfil$\crcr}}}
%
%
%
\let\sec@nt=\sec
\def\sec{\relax\ifmmode\let\n@xt=\sec@nt\else\let\n@xt\section\fi\n@xt}
\def\obsolete#1{\message{Macro \string #1 is obsolete.}}
\def\firstsec#1{\obsolete\firstsec \section{#1}}
\def\firstsubsec#1{\obsolete\firstsubsec \subsection{#1}}
\def\thispage#1{\obsolete\thispage \global\pagenumber=#1\frontpagefalse}
\def\thischapter#1{\obsolete\thischapter \global\chapternumber=#1}
\def\nextequation#1{\obsolete\nextequation \global\equanumber=#1
   \ifnum\the\equanumber>0 \global\advance\equanumber by 1 \fi}
\def\BOXITEM{\afterassigment\B@XITEM\setbox0=}
\def\B@XITEM{\par\hangindent\wd0 \noindent\box0 }
%

%
\catcode`@=12 
\message{ by V.K.}
\everyjob{\input myphyx }
\def\etal{{\it et al.}}
%
\vsize=8.75truein
\hsize=6.0truein
\voffset=-.3truein
\baselineskip=14pt
\overfullrule=0pt
%

\def\msun{M_\odot}
\def\lsun{L_\odot}

\def\pac{Paczy{\'n}ski}
\def\etal{{\it et al.}}

%
\bigskip\bigskip
\centerline{\bf  THE PARTICLE- AND ASTRO-PHYSICS OF DARK MATTER\foot{
Plenary talk presented at Snowmass 94 (Particle and Nuclear Astrophysics
and Cosmology in the Next Millennium, June 29 -- July 14, 1994, Snowmass,
Colorado.)}}
\bigskip
\centerline{KIM GRIEST}
\smallskip
\centerline{Physics Department}
\centerline{University of California, San Diego, La Jolla CA 92093}
\smallskip
\footline={\hss \tenrm \folio \hss}
\hbox{ }\smallskip
\centerline{\bf ABSTRACT}
We review some recent determinations of the amount of dark matter
on galactic, cluster, and large scales,
noting some puzzles and their possible resolutions.  We discuss the
interpretation of big bang nucleosynthesis for dark matter,
and then review the motivation for and basic physics
of several dark matter candidates, including
Machos, Wimps, axions, and neutrinos.
Finally, we discuss how the uncertainty
in the models of the Milky Way dark halo will affect the
dark matter detection experiments.

\bigskip\bigskip\bigskip
\unnumberedchapters
\baselineskip  12pt
\REF\zwicky{Zwicky, F. 1933. Helvetica Physica Acta, \bf 11, \rm 110.}
\REF\freeman{Freeman, K.C. 1970. \sl ApJ. \bf 160, \rm 811.}
\REF\ostriker{Ostriker, J.P. \etal, 1974. \sl ApJ. Lett. \bf 193, \rm L1.}
\REF\faber{Faber, S.M. \& Gallagher, J.S. 1979. \sl Ann. Rev. Astron. Astroph.
	\bf 17, \rm 135.}
\REF\sancisi{Sancisi, R. \& Van Albada, T.S. 1986, in \sl
	Dark Matter in the Universe, IAU Symp. No. 117, \rm Knapp, G. \&
	Kormendy, eds., p. 67 (Reidel, Dordrecht).}
\REF\trimble{Trimble, V. 1987. \sl Ann. Rev. Aston. Astroph. \bf 25, \rm 425.}
\REF\sadoulet{Primack, J.R., Sadoulet, B., \& Seckel, D. 1988. \sl
	Ann. Rev. Nucl. Part. Phys. \bf B38, \rm 751.}
\REF\walker{Walker, T.P., \etal, 1991. \sl ApJ. \bf 376, \rm 51.}
\REF\mond{Begeman, K.G., Broeils, A.H., \& Sanders, R.H. 1991. \sl
	M.N.R.A.S. \bf 249, \rm 532.}
\REF\Tremaine{Tremaine, S. \& Gunn, J. 1979. \sl Phys. Rev. Lett. \bf 42,
	\rm 407.}
\REF\spergel{Gerhard, O.E. \& Spergel, D.N. 1992, \sl ApJ. \bf 389, \rm L9.}
\REF\burrows{Burrows, A., Klein, D., \& Gandhi, R. 1992,
	\sl Phys. Rev. \bf D45, \rm 3361.}
\REF\fitch{Fich, M. \& Tremaine, S. 1991, \sl Ann. Rev. Astron. Astroph.
	\bf 29, \rm 409.}
\REF\vanbiber{Van Bibber, K., private communication.}
\REF\sikivie{Sikivie, P. 1983. \sl Phys. Rev. Lett. \bf 51, \rm 1415.}
\REF\axion{for example, Turner, M.S. 1990, \sl Physics Reports \bf 197, \rm 67;
	Raffelt, G.G. 1990, \sl Physics Reports \bf 198, \rm 1.}
\REF\paczynski{Paczynski, B. 1986. \sl ApJ. \bf 304, \rm 1.}
\REF\griest{Griest, K. 1991. \sl ApJ. \bf 366, \rm 412.}
\REF\gould{Gould, A. 1992, \sl ApJ. \bf 392, \rm 234.}
\REF\alcocketal{Alcock, C. \etal, 1991, in \sl Robotic Telescopes of the
	1990's, \rm ed. Filippenko, A.V. (ASP, San Fransisco).}
\REF\french{Moscoso, L., \etal, 1991, preprint Saclay-DPhPE91-81;
	Spiro, M., private communication.}
\REF\ogle{Udalski, A. \etal, 1992. \sl Acta Astronomica \bf 42, \rm 253;
	Paczynski, B., private communication.}
\REF\kolb{Kolb, E.W. \& Turner, M.S. 1990. \sl The Early Universe, \rm
	(Addison-Wesley, Redwood City, California).}
\REF\griestsadoulet{Griest, K. \& Sadoulet, B., in \sl Dark Matter in the
	Universe, \rm eds. Galeotti, P., \& Schramm, D.N. (Kluwer, Netherlands,
	1989).}
\REF\griestsilk{Griest, K. \& Silk, J. 1990. \sl Nature \bf 343, \rm 26.}
\REF\goodman{Goodman, W.E. \& Witten. E. 1985. \sl Phys. Rev. \bf D31, \rm
	3059.}
\REF\silk{Silk, J., Olive, K.A., \& Srednicki, M. 1985. \sl Phys. Rev. Lett.
	\bf 55, \rm 259.}
\REF\indirect{For example,
	Mori, M. \etal, 1991. \sl Phys. Lett. \bf B270, \rm 89;
	Sato, N. \etal, 1991. \sl Phys. Rev. \bf D44, \rm 2220;
	Barwick, S. \etal, 1992. \sl J. Phys. \bf G18, \rm 225.}
\REF\betty{Young, B.A, 1992, in Proceedings of the Third Intl. Conf. on
	Advanced Technology and Particle Physics, Borchi, E., \etal, eds.
	(Elsevier).}
\REF\ellishag{Ellis, J., Hagelin, J.S., Nanopoulos, D.V., Olive, K.A.,
	Srednicki, M. 1984. \sl Nucl. Phys. \bf B238, \rm 453.}
\REF\haber{Haber, H.E. \& Kane, G.L. 1985. \sl Physics Reports \bf 117, \rm
75.}
\REF\griestros{Griest, K. \& Roszkowski, L. 1992. \sl Phys. Rev.
	\bf D46, \rm 3309.}
\REF\shutt{Shutt, T. \sl et al. \rm 1992.  \sl Phys. Rev. Lett. \bf 69,
	\rm 3425; \sl ibid \rm 3531.}
\REF\kamion{Halzen, F., Stelzer, T., \& Kamionkowski, M. 1992, \sl Phys.
	Rev. \bf D45, \rm 4439.}
\REF\griestneut{Griest, K. 1988. \sl Phys. Rev. \bf D38, \rm 2357.}
\REF\minimalsugra{for example,
	Nojiri, M.M. 1991, \sl Phys. Lett. \bf B261, \rm 76;
	Ellis, J. \& Roszkowski, L. 1992, \sl Phys. Lett. \bf B283, \rm 252;
	Ross, G.G. \& Roberts, R.G. 1992, \sl Nucl. Phys. \bf B377, \rm 571;
	Drees, M \& Nojiri, M.M. 1992, \sl Phys. Rev. \bf D45, \rm 2482;
	Lopez, J.L., Nanopoulos, D.V., \& Yuan, K. 1992,
		\sl Nucl. Phys. \bf B370, \rm 445;
	Drees, M. \& Nojiri, M.M. 1992, preprint DESY-92-101;
	Roszkowski, L. 1992, private communication.}
\REF\dreestata{Drees, M. \& Tata, X. 1991, \sl Phys. Rev. \bf D43, \rm 2971.}
\REF\ibanez{Ibanez, L.E., \& Lust, D. 1992, \sl Nucl. Phys. \bf B382, \rm 305.}
\REF\randall{Hall, L.J., \& Randall, L. 1990, \sl Phys. Rev. Lett.
	\bf 65, \rm 2939.}
\REF\extendedsusy{For example, Flores, R.A., Olive, K.A., and Thomas, D. 1991,
	\sl Phys. Lett. \bf B245, \rm 514;
	Abel, S.A., Cottingham, W.N., \& Whittingham, I. 1990, \sl
	Phys. Lett. \bf B244, \rm 327;
	Flores, R.A., Olive, K.A., \& Thomas, D. 1991, \sl Phys. Lett.
	\bf B263, \rm 425.}
\centerline{\bf 1. Introduction}
\smallskip
It is remarkable that here, at the end of the 20th Century, when science
has produced the top quark, and measured tiny fluctuations in the microwave
background, we still don't know what the primary constituent of the
Universe is.  This is the stuff that dominates gravity on galactic scales,
and determines the ultimate fate of the Universe, but we have almost no idea
what it is.  It doesn't emit or absorb electromagnetic radiation at
any known wavelength, but it is ``seen" through its gravitational
effects on scales from tiny dwarf galaxies, to large spirals like the Milky
Way, to the largest scales yet observed.  The next speaker (Caldwell, these
proceedings) will describe exciting new developments in the possibility
of detecting this dark matter;  a topic which has progressed dramatically
in the past few years.
I will set the stage for his talk by reviewing recent developments in
the amount and location of dark matter,
mention some puzzles, and describe some of the more popular dark
matter candidates.

\bigskip
\centerline{\bf 2. Dark Matter Inventory}
\smallskip
Evidence for dark matter (DM) exists on many scales, and it is important to
keep in mind that the dark matter on different scales may be different --
the dark matter in
dwarf spirals may not be the dark matter which contributes $\Omega =1$;
in fact,
the $\Omega=1$ dark matter may not exist.  This consideration is
especially important when
discussing dark matter detection, since detection is done in the Milky Way,
and evidence for dark matter outside the Milky Way may not be relevant.

So, let me start with an inventory of dark matter in the Universe.
The quantity of dark matter on different scales is quoted using
$\Omega = \rho/\rho_{crit}$,
where $\rho$ is the density of some material averaged over the Universe,
and $\rho_{crit}$ is the critical density.  Most determinations of Omega
are made by measuring the mass-to-light ratio $\Upsilon$ of some system
and then multiplying this by the average luminosity
density of the Universe: $j_0 = 1.7 \pm .6 \times 10^8 h^{-1} \lsun/\msun$
(Binney and Tremaine 1987; Davis \& Huchra 1982; Kirshner \etal\ 1983).
Here $h = 0.4 - 1$ parameterizes
our uncertainty of the Hubble constant.
Note the factor of two uncertainty in this number, which
implies that
all determinations of $\Omega$ which use this method will be uncertain
by at least this amount.  In fact, almost all determinations of $\Omega$
do use this method;
the exceptions being determination of $\Omega_{baryon}$ from
big bang nucleosynthesis, and the large scale determinations from bulk flows.
For example, the mass-to-light ratio in the solar neighborhood
is $\Upsilon \approx 5$, giving $\Omega_{lum} = 0.003h^{-1} = 0.003 - 0.007$.
If the solar neighborhood is typical, the amount of material in
stars, dust and gas is far below the critical value.

\bigskip
\line{\bf 2.1 Spiral Galaxies \hfil}
\smallskip
The most robust evidence for dark matter comes from the rotation
curves of spiral galaxies.  Using 21 cm emission,
the velocities of clouds of neutral hydrogen can be measured as a function
of $r$, the distance from the center of the galaxy.  In almost all cases,
after a rise near $r=0$, the velocities remain constant out as far as
can be measured.  By Newton's law for circular motion $GM(r)/r^2 = v^2/r$,
this implies that the density drops like $r^{-2}$ at large radius and that
the mass $M(r) \propto r$ at large radii.  Once $r$ becomes greater
than the extent of the mass, one expects the velocities to drop $\propto
r^{-1/2}$, but this is never seen, implying that we do not know how large
the extended dark halos around spirals are.  For example, the rotation
curve of NGC3198 (Binney \& Tremaine 1987) implies $\Upsilon > 30h$, or
$\Omega_{halo} > 0.017$.  The large discrepancy between this number
and $\Omega_{lum}$ is seen in many external galaxies and is
the most robust evidence for dark matter.
This is fortunate since searches for dark matter can be made only in spiral
galaxies;  in fact only in our spiral, the Milky Way.  Unfortunately,
the rotation curve of the Milky Way is poorly constrained, which leads to
uncertainty in the amount of dark matter in our Galaxy.  However,
there is less secure evidence for substantial dark matter in our Galaxy.
By studying the motion of dwarf galaxies (especially Leo I at a distance
of 230 kpc) Zaritsky, \etal\ (1989) find a mass of the Milky Way of
$M_{MW} = 1.25^{+ 0.8}_{-0.3} \times 10^{12} \msun$, for $\Upsilon_{MW}
\approx 90$, and $\Omega_{MW} \approx 0.054 h^{-1}$  (assuming the Universe
is like the Milky Way).
There are a limited number of small satellite galaxies around the Milky
Way, so the uncertainty in this measurement is large.
However, for external galaxies Zaritsky (1992) used
a sample of 69 small satellite galaxies around 45 spirals similar to the
Milky Way to estimate the total mass for a ``typical" spiral.
He found that $M \approx 10^{12} \msun$ at 200 kpc from the center, implying
$\Omega_{spirals} \approx 0.087 h^{-1}$ out to this radius.
Even in this case, there is not strong evidence that the rotation speeds
drop, so there is no good upper limit to $\Omega_{spiral}$.
This number is similar to the number found by the Local Group Timing
method (e.g. Binney \& Tremaine 1987).

\vfil\eject
\line{\bf 2.2 Clusters of Galaxies \hfil}
\smallskip
Moving to larger scales, the methods of determining $\Omega$ become
less secure, but give larger values.  There is a great deal of new
evidence on dark matter in clusters of galaxies, coming from
gravitational lensing which Tyson (these proceedings) will discuss,
from X-ray gas temperatures, and from the motions of cluster member galaxies.
For example, consider the Coma cluster of around a thousand galaxies;
the cluster which Zwicky (1933)
used to first hypothesize that dark matter existed.
White \etal\ (1993) recently collated some of the data on the Coma
cluster, reporting separate measurements of the amount of mass in
stars, hot gas, and in total.  Within a radius of 1.5$h^{-1}$ Mpc, they
give
$$\eqalign{
M_{star} =& 1.0 \pm 0.2 \times 10^{13} h^{-1} \msun \cr
M_{gas} =& 5.4 \pm 1 \times 10^{13} h^{-5/4} \msun \cr
M_{total} =& 5.7 - 11 \times 10^{14} h^{-1} \msun, \cr}
$$
where the total mass is estimated in two completely different ways.
The first method is a refinement of Zwicky's method of
using the radial velocities of the member galaxies, and the assumption
of virialization to gauge the depth of the gravitational potential well.
The second method makes use of the ROSAT X-ray maps and the assumption of
a constant temperature equilibrium to get the same information.
Remarkably the two methods give the same mass within errors.
Thus with a mass-to-light ratio of $\Upsilon = 330 - 620 \msun/\lsun$,
one finds $\Omega= 0.2 - 0.4$, if the inner 1.5 Mpc of
Coma is representative of the Universe as a whole.

There is, however, a very disconcerting fact about the above numbers.
As pointed out by White, \etal\ (1993),
$$
{M_{baryon} \over M_{total}} > 0.009 + 0.05 h^{-3/2}.
$$
Now the Coma cluster is large enough that one might expect its baryon
to dark matter ratio to be the Universal value,
($\Omega_{baryon}/\Omega_{total} = M_{baryon}/M_{total}$), and in fact
White, \etal\ argue that this is the case.  Then the inequality above
should apply to the entire Universe.  But, as we discuss later,
big bang nucleosynthesis limits $\Omega_{baryon} < 0.015h^{-2}$.
If $\Omega_{total} = 1$, as many would like,
the two inequalities are in quite strong disagreement for any value of $h$!
So this is a big puzzle.  The conclusions of White, \etal, are that
either $\Omega$ is not unity, or that big bang nucleosynthesis is not
working.  As I think Joel Primack (these proceedings)
and Tony Tyson (these proceedings) will discuss, there are other possible
explanations.  Notably that measurements of the the total mass in clusters
by weak or strong gravitational lensing tend to give twice as much
total mass as the X-ray and virial methods, and that mass and velocity
bias may mean that clusters are not so representative of the Universe
as a whole.

\vfil\eject
\line{\bf 2.3 Large Scale Flows \hfil}
\smallskip
Next, we turn to some of the most interesting new results on
the amount of dark matter.   One would really like to measure
the amount of dark matter on the largest possible scales so that
one can be sure the sample is representative of the entire Universe.
Within the past several years a host of related methods have been
tried, and while they initially gave uncertain results, the most
recent determinations do impressively well.  These methods
have the advantage stated above, but the disadvantage that they
depend upon assumptions about galaxy formation;  which means they
depend upon gravitational instability theory, biasing, etc.
Also, the errors in these results are still large and the
calculations are complicated, but they do have the promise to answer
the fundamental question of how large $\Omega$ is.
They also tend to give values of
$\Omega_{total}$ near unity!

The simplest example comes from
the observation that the local group of galaxies moves
at $627 \pm 22$ km/sec with respect to the Cosmic
Microwave Background (CMB) (measured from the amplitude of
the CMB dipole).
If this motion comes from gravity, then the direction of the motion should
line up with the direction where there is an excess of mass, and the
velocity should be determined by the size of this excess.
Thus, taking into account the expansion of the Universe, one has
$$
v \propto \Omega^{0.6} {\delta\rho\over\rho} = {\Omega^{0.6} \over b}
{\delta n\over n},
$$
where the bias factor $b$ has slipped in because observers use
the excess in galaxy number counts $\delta n/n$ to estimate
the excess in mass density $\delta\rho/\rho$.
So, using galaxy counts from the IRAS satellite survey,
Yahil \etal\ (1986) find
that the direction of the $\delta n/n$ excess agrees with the direction
of the velocity vector to within $\sim 20^0$, and that
$$
{\Omega^{0.6}\over b} = 0.9 \pm 0.2.
$$
Thus with the very conservative limit $b>0.5$, one has
$\Omega>0.2$, and with
the reasonable limit $b>1$, one finds $\Omega>0.5$.
So there is now observational evidence that $\Omega$ is near
the theoretically favored value of unity.  However, for the method
above to be reliable, one must measure $\delta n/n$
on very large scales to ensure that convergence has been reached.
I don't think this has yet been convincingly demonstrated.

The above technique is only one of a host of methods used to
determine $\Omega$ on large scales.  Especially
promising is the detailed
comparison of the peculiar velocities of many galaxies to
the detailed maps of $\delta n/n$.  This should not only determine
$\Omega$, but serve as a stringent test on the theory that
large scale structure is formed by gravitational instability.
The trick is to get the peculiar velocities, which means subtracting
the much larger Hubble flow velocity.
Since the redshift measurements gives only
the radial component of velocity
it seemed difficult to get complete information,
but Bertshinger and Dekel (1989) proposed an ingenious method, in which
one assumes that the velocity field is curl free, and proceeded to
reconstruct the entire three dimensional field.  They use
$$
\bigtriangledown \cdot {\bf v} = -{\Omega^{0.6} \over b}{\delta n\over n},
$$
and solve for $\Omega^{0.6}/b$.  They find $0.3 < \Omega < 1$ for
reasonable choices of $b$.
Especially notable is that the detailed
$\delta n/n$ maps agree remarkably well with
the reconstructed velocity fields, and I interpret this as
evidence that gravitational instability is the most likely cause
of the structure.
In fact, this is now the story for most of the large scale flow methods.
They find reasonable to excellent agreement with the theory,
and predict a large value of $\Omega$.
A very nice table of values and a review of many of these methods
can be found in Dekel (1994).

\bigskip
\line{\bf 2.4 Big Bang Nucleosynthesis \hfil}
\smallskip
Next, I turn to a very important and recently controversial
ingredient in the dark matter story; namely the predictions
of big bang nucleosynthesis.  The standard story has been
(Smith, Kawano, Malaney 1993; Walker, \etal\ 1991) that
to get agreement with the measured abundances of helium, deuterium,
and lithium, the baryonic content of the Universe had to
be between $0.01 \leq \Omega_{baryon}h^2 \leq 0.015$.
Given the large uncertainty in $h$ this meant
$0.01 \leq \Omega_{baryon} \leq 0.1$.  This value is inconsistent
with having a total $\Omega=1$ in baryons, and so, together with
the predilection for $\Omega_{total}=1$, is the main motivation
for postulating non-baryonic (elementary particle) dark matter.
The majority of dark matter searches are targeted towards these
non-baryonic particles.  The lower limit of this range is actually
{\it above} the abundance of known stars, gas, etc., and so
there also seems to be evidence for substantial {\it baryonic}
dark matter as well.

Much recent excitement was caused in this field when
Songaila, \etal\ (1994) reported a possible detection of deuterium
in a Lyman limit cloud at a redshift of 2.9 (via absorption
of light emitted by an even more distant quasar).
The controversy was caused by the extremely large value
D/H $\approx 2.4 \times 10^{-4}$, when measurements from the
interstellar medium are in the range $1.9 \times 10^{-5} < {\rm D/H} <
6.8 \times 10^{-5}$ (Steigman \& Tosi 1992).
If the new measurement is true, the
limits from big bang nucleosynthesis would move to roughly to
$0.004 < \Omega h^2 < 0.0068$, or
$0.006 \leq \Omega \leq 0.035$, substantially lower than the current
limits.
In this case, most of the baryonic material may well
be known, and models of structure formation which require substantial
baryonic content may be in trouble.  However, it is too soon
to tell whether the measurement will stand up.  Several groups
are making these measurements and hopefully soon there will
be either confirmation or refutation of this result.

\bigskip
\line{\bf 2.5 Distribution of Dark Matter in the Milky Way \hfil}
\smallskip
While we don't have a clue as to what the dark matter (DM) is,
we have a reasonable idea as to how much of it there is in the Galaxy,
how it is distributed, and how fast it is moving.
This information comes from the rotation curve of the Milky Way,
and is crucial to all the direct searches for dark matter.
If we say that the rotation curve of the Milky Way is constant at about
$v_c = 220$ km/sec out to as far as it is measured, then we know that
the density must drop as $r^{-2}$ at large distances.
This velocity also sets the scale for the depth of the potential well
and says that the dark matter must also move with velocities in this range.
Assuming a spherical and isotropic velocity distribution is common,
and a usual parameterization is
$$
\rho({\bf r}) = \rho_0 {a^2 + r_0^2\over a^2 + r^2},
$$
where $r_0 \approx 8.5$ kpc is the distance of the Sun from the galactic
center, $a$ is the core radius of the halo, and $\rho_0 \approx 0.3\
{\rm GeV~cm}^{-3}$ is the density of dark matter near the Sun.
Also, a typical velocity distribution is
$$
f(v) d^3v = {e^{-v^2/v_c^2} \over \pi^{3/2}v_c^3} d^3v.
$$
It should be noted that the specifics of the above models are not
very secure.  For example, it is quite possible that the halo of
our Galaxy is flattened into an ellipsoid, and there may be a component of
the halo velocity which is rotational and not isotropic.
Also, some (or even much) of the rotation curve of the Milky Way
at the solar radius could be due to the stellar disk.  Canonical models of the
disk have the disk contributing about half the rotation velocity,
but larger disks have been envisioned.  Recent microlensing results
may be indicative of a larger disk as well.  I will talk about this later
if there is time.

Finally, other important points about our Galaxy's geography include
the fact that the nearest two galaxies are the LMC and SMC, located
at a distance of 50 kpc and 60kpc respectively, that the halo of the
Milky Way is thought to extend out at least this far, and that
the bulge of the Milky Way is a concentration of stars in the center
of our Galaxy (8.5 kpc away) with a size of about 1 kpc.

\bigskip
\centerline{\bf 3. Dark Matter Candidates}
\smallskip
There is no shortage of ideas as to what the dark matter could be.
In fact, the problem is the opposite.  Serious candidates
have been proposed with masses ranging
from $10^{-5}$ eV = $1.8 \times 10^{-41}$ kg $= 9 \times 10^{-72} \msun$
(axions) up to $10^4 \msun$ black holes.  That's a range of masses
of over 75 orders of magnitude!  It should be clear that no one
search technique could be used for all dark matter candidates.

Even finding a consistent categorization scheme is difficult, so
we will try a few.  First, as discussed above, is the baryonic
vs non-baryonic distinction.  The main baryonic candidates
are the Massive Compact Halo Object (Macho) class of candidates.
These include
brown dwarf stars, jupiters, and 100 $\msun$ black holes.
Brown dwarfs are balls of H and He with masses below 0.08 $\msun$,
so they never begin nuclear fusion of hydrogen.
Jupiters are similar but with masses near 0.001 $\msun$.
Black holes with masses near 100 $\msun$
could be the remnants of an early generation of stars
which were massive enough so that not many heavy elements were
dispersed when they went supernova.  Other, less popular, baryonic
possibilities
include fractal or specially conditioned clouds of neutral hydrogen.
The non-baryonic candidates are basically elementary particles which
are either not yet discovered or have non-standard properties.
Outside the baryonic/non-baryonic categories
are two other possibilities which
don't get much attention, but which I think should be kept in mind
until the nature of the dark matter is discovered.  The first is non-Newtonian
gravity.  See Begeman \etal\ (1991) for a provocative discussion of this
possibility; but watch for results from gravitational lensing which
may place very strong constraints.  I won't talk about this possibility here.
Second, we shouldn't ignore the ``none-of-the-above" possibility
which has several times surprised the physics/astronomy community in the past.

Among the non-baryonic candidates there are several classes of
particles which are distinguished by how they came to exist in large quantity
during the Early Universe,
and also how they are most easily detected.
The axion (Section 3.6) is motivated as a possible solution
to the strong CP problem and is in a class by itself.  The largest
class is the Weakly Interacting Massive Particle (Wimp) class (Section 3.4),
which consists of literally hundreds of suggested particles.
The most popular of these Wimps
is the neutralino from supersymmetry (Section 3.5).
Finally, if the tau or muon neutrino had a mass in the 5 eV to 100 eV
range, it could make up all or much of the dark matter (Section 3.3).

Another important categorization scheme is the ``hot" vs ``cold"
classification.  A dark matter candidate is called ``hot" if it was moving
at relativistic speeds at the time galaxies could just start to form
(when the horizon first contained about $10^{12}\msun$).  It is called
``cold" if it was moving non-relativistically at that time.
This categorization
has important ramifications for structure formation, and there is
a chance of determining whether the dark matter is hot or cold
from studies of galaxy formation.  Hot dark matter cannot cluster
on galaxy scales until it has cooled to non-relativistic speeds,
and so gives rise to a considerably different primordial fluctuation
spectrum.
Of the above candidates only the light neutrinos would be hot;  all the
others would be cold.

\bigskip
\line{\bf 3.2 Machos \hfil}
\smallskip
Probably the most exciting development in the dark matter story
is the detection of Machos by three separate groups
(Alcock, \etal\ 1993; Aubourg, \etal\ 1993; Udalski, \etal\ 1993).
All three groups monitored millions of
stars, either in the LMC or in the galactic bulge, for signs
of gravitational microlensing, and all three groups seem to have found it.
It is still not clear whether the objects which are causing the microlensing
are numerous enough to make up the Milky Way dark matter, or even
whether they are in the galactic halo or disk.  Whether they are
light enough to qualify as bona fide brown dwarfs, or whether they
are just faint stars is also not yet known.
See David Caldwell's talk (these proceedings)
for a review of the current status of these important experiments.

However, from the more general point of view, I'd like to note
that these experiments very well may have the capability to
give a definitive answer to the question of whether the dark matter
in our Galaxy is baryonic.
The microlensing searches are probably sensitive to
any objects in the range $\sim 10^{-8}\msun < m < 10^3 \msun$,
just the range in which such objects are theoretically allowed to
exist.  Objects made purely of H and He with masses
less than $\sim 10^{-9} -10^{-7}\msun$ are expected to evaporate
due to the microwave background in less than a Hubble time, while
objects with masses greater than $\sim 10^3\msun$ would
have disrupted known globular clusters.

\bigskip
\line{\bf 3.3 Light Neutrinos \hfil}
\smallskip
Hot dark matter was out of favor for several years, but
has come back into style in a big way.
Neutrino dark matter was unpopular for two main reasons.  First, by Jeans
theorem, the current
phase space density of any dissipationless particle should be less than
or equal to
its phase space density at decoupling.  Tremaine \& Gunn (1979),
showed that this meant very light neutrinos could not be packed
endlessly into dwarf galaxies.  Gerhard \& Spergel (1992) applied
this result to the dwarf galaxy Ursa Minor, which is known to have
a great deal of dark matter, and very conservatively
found that if all the dark matter in Ursa Minor were neutrinos,
the neutrinos would have to have masses greater than 81 eV.
Since $\Omega_\nu h^2 = m_\nu /90\ {\rm eV}$,
and $\Omega_\nu h^2 <1$, the maximum mass which
neutrino dark matter could have is around 90 eV, with values of around
30 eV being favored.
So the dark matter in dwarf galaxies such as Ursa Minor and Draco
probably cannot be light neutrinos.  Thus, one would need at least two
types of dark matter to have hot dark matter play an important role.
By simplicity, this led to neutrino dark matter being disfavored.

The other reason neutrino dark matter became
less popular was that galaxy formation with pure hot dark matter
and a Zeldovich
perturbation spectrum just doesn't match the observations.

While the arguments above are still valid, several recent developments
have changed the attitude towards them.  First, it is now known
that galaxy formation with pure cold dark matter and
a Zeldovich perturbation spectrum also doesn't match the observations.
So some additional ingredient is needed.  Second, one of the few models
which does match the observations is a mixed dark matter Universe
with $\Omega_{cold}=.7$, $\Omega_\nu=0.25$, and $\Omega_{baryon}=0.05$.
Thus, since it seems we may need two types of dark matter anyway,
the objection that the dark matter in dwarf galaxies cannot all be neutrinos
becomes less important.  We may well have a tau or muon neutrino
with a mass near 6 eV.  In these mixed models, the cold dark matter
dominates, and so the ongoing searches for the other dark matter
candidates are hardly affected.

See talks by Joel Primack and David Caldwell (these proceedings) for
more discussion of these models and attempts at
detection of neutrino dark matter by measuring the neutrino mass.

\bigskip
\line{\bf 3.4 Thermal Relics as Dark Matter (Wimps) \hfil}
\smallskip
Among the particle dark matter candidates an important distinction
is whether the particles were created thermally in the Early Universe,
or whether they were created non-thermally in a phase transition.
Thermal and non-thermal relics have a different relationship between
their relic abundance $\Omega$ and their properties
such as mass and couplings, so the distinction is especially important
for dark matter detection efforts.
For example, the Wimp class of particles can be defined as those particles
which are created thermally, while dark matter
axions come mostly from non-thermal processes.

In thermal creation one imagines that early on, when the Universe
was at very high temperature, thermal equilibrium obtained, and
the number density of Wimps (or any other particle species)
was roughly equal to the number density of photons.
As the Universe cooled the number of Wimps and photons would decrease together
as long as the temperature remained higher than the Wimp mass.
When the temperature finally dropped below the Wimp mass, creation
of Wimps would require being on the tail of the thermal distribution,
so in equilibrium, the number density of Wimps would drop
exponentially $ \propto \exp(-m_{Wimp}/T)$.  If equilibrium were maintained
until today there would be very few Wimps left, but at some point
the Wimp density would drop low enough that
the probability of one Wimp finding another
to annihilate would become small.  (Remember we must assume that an individual
Wimp is stable if it is to become the dark matter.)
The Wimp number density would ``freeze-out" at this point
and we would be left
with a substantial number of Wimps today.  Detailed evolution
of the Boltzmann equation can be done for an accurate prediction, but
roughly
$$
\Omega_{Wimp} \approx {10^{-26} {\rm cm}^3 {\rm s}^{-1}
\over \VEV{\sigma v}},
$$
where $\VEV{\sigma v}$ is the thermally averaged cross section
for two Wimps to annihilate into ordinary particles.
The remarkable fact is that for $\Omega \approx 1$, as required by
the dark matter problem, the annihilation cross section
$\left<\sigma v\right>$ for any thermally created particle
turns out to be just
what would be predicted for particles with electroweak scale interactions.
Thus the name ``Wimp".  There are several theoretical problems with the
Standard Model of particle physics which are solved by new electroweak
scale physics such as supersymmetry.  Thus these
theoretical problems may be clues that the dark
matter does indeed consist of Wimps.  Said another way, any stable
particle which annihilates with an electroweak scale cross section
is bound to contribute to the dark matter of the Universe.  It is
interesting that theories such as supersymmetry, invented for entirely
different reasons, typically predict just such a particle.

The fact that thermally created dark matter has weak scale interactions
also means that it may be within reach of accelerators such as LEP at CERN,
and CDF at Fermilab.  After all these accelerators were built
precisely to probe
the electroweak scale.  Thus many accelerator searches for exotic particles
are also searches for the dark matter of the Universe.
Also, due to the weak scale interactions, Wimp-nuclear
interaction rates are within reach of many direct and indirect
detection methods.  Caldwell (these proceedings) will review these.

\bigskip
\line{\bf 3.5 Example of the Neutralino \hfil}
\smallskip
The most popular of the Wimp candidates is the neutralino from supersymmetry.
Supersymmetry seems to be a necessary ingredient of any theory which
which consistently combines general relativity with the Standard Model
of particle physics.

When Dirac attempted to combine special relativity with
quantum mechanics, he came up with the Dirac equation and discovered
(to his chagrin) that it contained a new CPT symmetry which required
the existence of a CPT partner, or anti-particle for every
known particle.  Dirac's initial hopes that the electron might be
the anti-particle of the proton were soon dashed, but
the discovery of the positron vindicated Dirac's theory.
Today this ``doubling" of the number of particles by CPT symmetry is taken
for granted to such an extent that anti-particles are not listed
in the particle data book.

It is interesting that attempts to combine general relativity with
quantum field theory seem to require supersymmetry in a similar way.
Here the symmetry relates integral spin particles to half integral spin
superpartners and vice versa.
Analogous initial hopes that the photon might be the superpartner of the
neutrino were soon dashed, and now, if supersymmetry exists, it is
known that there would be a doubling of the number of known particles.
While there has been no quickly following discovery
of supersymmetric particles, the hypothetical
particles have been named and thoroughly studied.  Intense searches
for these particles are taking place at all the major accelerators.
Examples of superpartners are the spin 1/2 photino, Higgsino,
Z-ino, and W-ino (superpartners of the photon, Higgs, Z, and W), and
the spin 0 squark and selectron (partners of the quark and electron).

There are actually several other strong motivations for the existence
of supersymmetry, among which, the stabilization of the electroweak
scale (hierarchy problem), and the remarkable coupling constant
unification which occurs in supersymmetric grand unified models,
stand out.
The interest for dark matter, however,  arises because typically there is a
conserved multiplicative R-parity, which means that the lightest
supersymmetric particle (LSP) is stable.
The most likely LSP is the neutralino, which is a
linear combination of the photino, Z-ino, and Higgsino.
The precise components of the combination are determined by the
parameters of the underlying supersymmetric model, and there are
many of these parameters.
Typically, the minimal supersymmetric model (MSSM) is
considered, which is the supersymmetric extension of the
Standard Model with the minimal number of new particles.
While supersymmetry relates many couplings and masses,
there are still dozens of free parameters in the most general MSSM, so
many times a restricted set of these parameters is used which derives
from assuming a simple grand unification scheme.
The particle accelerators are searching this parameter space for supersymmetry,
but have turned up no new particles.
It is interesting is that in much of this parameter space
calculation of the relic abundance
predicts $\Omega \approx 1$ in neutralinos.
For example a recent calculation by Drees and Nojiri (1993)
(as programmed by Jungman, \etal\ 1994) shows this clearly.

These supersymmetric models are also within
reach of the new generation of direct
and indirect Wimp detection experiments as
Caldwell (these proceedings) will review.

\bigskip
\line{\bf 3.6 Axions \hfil}
\smallskip
\def\tbar{{\bar\theta}}
The best example of a non-thermal particle dark matter candidate
is the axion.  Actually thermal axions are produced in the standard
way, but if such axions existed in such numbers as to make up the dark
matter, they would have lifetimes too short to still be around in
quantity.
However, there is another, more important, production mechanism
for axions in the early Universe.

The axion arises because the QCD Lagrangian contains a term
$$ L \supset {\tbar g^2 \over 32 \pi^2} G \tilde G,
$$
where $G$ is the gluon field strength.
This term predicts an electric dipole moment of the neutron of
$d_n \approx 5 \times 10^{-16} \tbar$.
Experimentally, however, the neutron dipole moment
$d_n < 10^{-25}$, which means $\tbar  < 10^{-10}$.
The question becomes why does this $\tbar$ parameter have such
a small value, when it naturally would have a value near unity?
This is the strong CP problem, and one way to resolve this
problem is to introduce a new Peccei-Quinn symmetry which predicts
a new particle named the axion.  The P-Q symmetry forces $\tbar=0$
at low temperatures today, but in the early Universe,
the axion field was free to roll around the bottom
of its Mexican hat potential.  The axion field motion in
the angular direction is called $\theta$, and since the curvature of
the potential in this direction is zero, the axion at high temperatures
was massless.  However, when the temperature of the Universe cooled
below a few hundred MeV (QCD energy scale), the axion potential
``tilts" due to QCD instanton effects, and the axion begins to oscillate
around the minimum, like a marble in the rim of a tilted Mexican hat.
The minimum of the potential forces the average $\tbar$ to zero,
solving the strong CP problem, and the curvature of the potential
means the axion now has a mass.  There is no damping mechanism for
the axion oscillations, so the energy density which goes into oscillation
remains until today as a coherent axion field condensate filling the
Universe.  This is a zero momentum condensate and so constitutes
cold dark matter.  One can identify this energy density as a bunch
of axion particles, which later can become the dark matter
in halos of galaxies.  The relic energy density $\Omega$ is thus related
to the tilt of the potential, which in turn is related to the axion
mass, a free parameter of the model.  If the axion mass
$m_a \approx 10^{-5}$ eV, then $\Omega_a \approx 1$.
This rather unusual story is still probably the most elegant
solution to the strong CP problem, and several groups are
mounting laboratory searches for the coherent axions which may make up
the major component of mass in the Galaxy.
Again Caldwell (these proceedings) will review these experiments.

\bigskip
\centerline{\bf 4. Uncertainties in the Milky Way Dark Matter}
\smallskip
Finally, let me turn to a subject which has not received a lot
of attention, but which I think will get increasing attention
now that one of the dark matter searches (Macho) is returning
positive results. That is, the uncertainties in the results of dark
matter searches due to uncertainties in the model of
Milky Way dark matter halo, where by model I mean the
density and velocity distributions of the dark matter.
This uncertainty affects
Wimp, axion, and Macho searches, and so should be given some attention.
For example, axion detection rates are proportional to the
local density of dark matter, which can vary greatly in different galactic
models, while Wimp rates are proportional to this as well as to
the average Wimp velocity.
Macho microlensing rates, being averages over the
line-of-sight to the stellar source are more complicated functions
of all the parameters in the halo model.

Much of the model dependence comes because the parameters of Milky
Way (and especially of the dark halo) are not well known.
For example, the local circular velocity has been measured at values
in the range $190\ {\rm km/s} < v_c < 250\ {\rm km/s}$.
Estimates of the distance of the Sun from the galactic center range
between $7 < r_0 < 9$ kpc. And the halo core radius
has been estimated at between $2 < a < 10$ kpc.
In addition, the halo may not be spherical, but may be flattened into
an ellipsoidal configuration, and the rotation curve may be gently
rising or falling (by about 15\%).

To investigate this model dependence (Alcock, \etal, 1994a)
one can use some nice self consistent
models of galactic halos due to Evans (1994).  For example,
one can calculate the microlensing rate towards the LMC, SMC, and bulge
using these models and see how the predicted rates change
as the parameters vary within their observationally allowed ranges.
Allowing the halo to be flattened up to a very reasonable axis ratio
of three-to-one, one finds a variation in the predicted microlensing rate of
more than a factor of ten.  Since the purpose of the microlensing
experiments is to measure the fraction of the dark halo made
of Machos, and the connection between the measurement and
fraction requires a model of the halo, one sees that this large
model uncertainty will translate into an uncertainty in the dark matter
fraction.  Thus it may be difficult for the Macho searches to place
strong constraints on the amount of non-baryonic dark matter in the halo.

However, there is an additional, and perhaps even larger,
uncertainty which comes about because
the mass of the stellar disk is not well known.
In the above derivations of the allowed halo parameters a ``canonical"
exponential stellar disk was used.  This says that about half
the rotation curve of the Galaxy comes from the halo, and about
half from the stellar disk  (Evans \& Jijina 1994).
However, if, as has been suggested
for many years, the disk is actually twice as
massive (Oort 1960; Bahcall 1984), then
very little need would exist for a dark halo component at the solar
radius of 8.5 kpc. (At large $r$ there would still be a
need for halo dark matter).
This might mean that the Wimp and axion searches would not
have much dark matter locally to look for.
The Macho searches are hoping to measure the amount of material in
the disk and so resolve this part of the problem
(Griest, \etal\ 1991;  \pac, 1991; Alcock, \etal\ 1994b; Udalski, \etal\ 1994).

The hope is that the microlensing searches can also resolve
the halo model uncertainties (if the halo consists of machos!).
The idea is to look in several different lines-of-sight (LMC, SMC, M31,
and bulge) and find quantities which differ enough among the different
models to distinguish them.

\bigskip
\centerline{\bf 5. Conclusions}
\smallskip
Rapid progress is being made in the dark matter question, but
as usual with new results, more questions are being opened than closed.

We thank Andrew Gould, George Fuller, Joel Primack, and members
of the MACHO collaboration for valuable help.
K.G. acknowledges support from a DoE OJI Award,
the NSF Center for Particle
Astrophysics, and the Alfred P. Sloan foundation.

\bigskip
\line{\bf References \hfil}
\medskip
\def\refind{\noindent\hangindent=1.5pc\hangafter=1}

\refind
Alcock, C. \etal\ 1993, Nature, 365, 621

\refind
Alcock, C. \etal\ 1994a, submitted to ApJ

\refind
Alcock, C. \etal\ 1994b, submitted to ApJ

\refind
Ashman, K. 1992, PASP, 104, 1109

\refind
Aubourg, E. \etal\ 1993, Nature, 365, 623

\refind
Bahcall, J.N. 1984, ApJ, 287, 926

%
\refind
Begeman, K.G, Broeils, A.H, \& Sander, R.H. 1991, MNRAS, 249, 523

\refind
Binney, J. \& Tremaine, S. 1987, Galactic Dynamics (Princeton University
Press, Princeton)

\refind
Bertshinger \& Dekel 1989, ApJ Lett, 336, L5

%
\refind
Davis, M. \& Huchra, J. 1982, ApJ, 254, 437

\refind
Dekel, A. 1994, to appear in ARAA 32.

\refind
Drees, M. \& Nojiri, M.M., 1993, Phys. Rev. D48, 3483

\refind
Evans, N.W. 1994 MNRAS, 267, 333

\refind
Evans, N.W., \& Jijina, J. 1994, MNRAS 267, L21

%
%
\refind
Gerhard, O.E. \& Spergel, D.N. 1992, ApJ. 389, L9

\refind
Griest, K. 1991, ApJ, 366, 412

\refind
Griest, K. \etal\ 1991, ApJ Lett, 372, L79

\refind
Jungman, J., Griest, K, \& Kamionkowski, M. 1994, in progress

\refind
Kirshner, R.D., \etal\ 1983, AJ, 88, 1285

\refind
Oort, J.H. 1960, Bull. Astr. Inst. Netherlands, 6, 249

\refind
\pac, B. 1986, ApJ, 304, 1

\refind
\pac, B. 1991, ApJ Lett, 371, L63

\refind
Primack, J.~R., Seckel, D., \& Sadoulet, B. 1988, Ann. Rev. Nucl.
Part. Sci., 38, 751

\refind
Smith, M.S., Kawano, L.H., \& Malaney, R.A. 1993, ApJ Supp, 85,219

\refind
Songaila, A., \etal\ 1993, Nature, 368, 599

\refind
Steigman, G \& Tosi, M. 1992, ApJ, 401, 150

\refind
Tremaine, S. \& Gunn, J. 1979. Phys. Rev. Lett., 42, 407

\refind
Udalski \etal\ 1993, Acta Astron, 43, 289

\refind
Udalski \etal\ 1994, preprint

\refind
Walker, T.P. \etal\ 1991, ApJ, 376, 51

\refind
White, S.D.M., \etal\ 1993, Nature, 366, 429

\refind
Yahil, A., Walker, X., \& Rowan-Robinson, M. 1986, ApJ Lett, 301, L1

\refind
Zaritsky, D. 1992, PASP, 394, 1

\refind
Zaritsky, D. \etal, 1989, ApJ, 345, 759

\refind
Zwicky, F., 1933. Helvetica Physica Acta, 11, 110
\vfill
\end